\documentclass[preprint,prd,aps,showpacs,showkeys,nofootinbib]{revtex4}
\usepackage{amssymb}
\usepackage{}
\usepackage{amsmath}
\usepackage{graphicx}
\usepackage{float}

\begin{document}

\preprint{}

\title{The LFV decays $B^0_{d,s}\rightarrow e\mu(e\tau,\mu\tau)$ with one neutral singlet scalar}

\author{Ke-Sheng Sun$^a$\footnote{sunkesheng@126.com or sunkesheng@mail.dlut.edu.cn}, Xiu-Yi Yang$^{b}$ \footnote{yxyruxi@163.com} }

\affiliation{$^a$ Department of Physics, Baoding University, Baoding, 071000,China\\
$^b$ College of Science, University of Science and Technology Liaoning, Anshan, 114051, China}

\begin{abstract}
Taking account of the constraint from radiative two body decays $l_i\rightarrow l_j\gamma$, we investigate the Lepton Flavor Violation decays $B^0_q\rightarrow\bar{l_l}l_k$ in the framework of the minimal extension of the Standard Model with one neutral singlet scalar. The couplings $C_{e\mu}$, $C_{e\tau}$ and $C_{\mu\tau}$ between the different generation leptons and scalar $S^0$ are constrained by the current bounds of $l_i\rightarrow l_j\gamma$. The numerical results show that the theoretical prediction of $B^0_q\rightarrow\bar{l_l}l_k$ strongly depend on the couplings $C_{qb}$ ($q=d\;or\; s$) between down type quarks and new scalar. The contributions from couplings $C_{uc}$, $C_{ut}$ and $C_{ct}$ between up type quark and new scalar are less dominant.

\end{abstract}

\keywords{B meson; lepton flavor violation } \pacs{13.20.He}

\maketitle

\section{Introduction}
\indent\indent

Rare decays are of great importance in searching for New Physics (NP) beyond the Standard
Model (SM), and the Lepton Flavor Violating (LFV) decays are particularly appealing cause
they are suppressed in the Standard Model (SM), and their detection would be a manifest
signal of NP. The LFV decays are discussed in various NP models, such as grand unified models \cite{GUT1,GUT2,GUT3},
supersymmetric models with and without R-parity \cite{SUSY1,SUSY2}, models with heavy sterile fermions \cite{sterile1,sterile2,sterile3,sterile4} and extra $Z^0$ boson \cite{Zp1,Zp2},left-right symmetry models \cite{LR1,LR2} etc.
Most of the current experimental focuses in searching for the LFV decays are lepton decays, $l_i\rightarrow l_j\gamma$, $l_i\rightarrow 3l_j$ and the $\mu-e$ conversion in nuclei. The LFV decays of hadrons are of great importance as well as the leptonic decays \cite{sun,dong}.

\begin{table}[h]
\caption{Current limits of LFV decays of $B^0_{s,d}$. }
\begin{tabular}{@{}cccc@{}} \toprule
Decay&Bound&Decay&Bound\\
\colrule
$B^0_{d}\rightarrow e^{\pm}\mu^{\mp}$&$< 2.8\times 10^{-9}$&$B^0_{s}\rightarrow e^{\pm}\mu^{\mp}$&$<1.1\times 10^{-8}$\\
$B^0_{d}\rightarrow e^{\pm}\tau^{\mp}$&$< 2.8\times 10^{-5}$&$B^0_{s}\rightarrow e^{\pm}\tau^{\mp}$&-\\
$B^0_{d}\rightarrow \mu^{\pm}\tau^{\mp}$&$< 2.2\times 10^{-5}$&$B^0_{s}\rightarrow \mu^{\pm}\tau^{\mp}$&-\\ \botrule
\end{tabular}
\label{tab1}
\end{table}
In literature, the LFV processes are associated with the lepton nonuniversality effect in semileptonic decays and $b\rightarrow sll$ transitions. These processes have been investigated in various models beyond the SM, such as supersymmetric models \cite{susy}, models extended with extra gauge $Z^{'}$ boson \cite{Zp3}, heavy singlet Dirac neutrinos \cite{HDn} or leptoquarks \cite{Lq,Lq1} and the Pati-Salam model \cite{PS}. Current experimental upper bounds on LFV decays of $B^0_{s,d}$ are listed in TABLE. \ref{tab1} \cite{PDG}. The experimental data on the LFV decays $B^0_{s}\rightarrow e^{\pm}\tau^{\mp}$ and $B^0_{s}\rightarrow \mu^{\pm}\tau^{\mp}$ are absent.
The theoretical prediction on the branching fractions of $B^0_{d,s}\rightarrow e^{\pm}\mu^{\mp}$ in these models can be greatly enhanced, even up to $10^{-11}$, which are very promising detected in near future. In literatures, the branching ratios of $B^0_{d,s}\rightarrow e^{\pm}\tau^{\mp}$ and $B^0_{d,s}\rightarrow \mu^{\pm}\tau^{\mp}$ can also be enhanced close to $B^0_{d,s}\rightarrow e^{\pm}\mu^{\mp}$ \cite{Lq,Lq1}. Recently, based on a sample of proton-proton collision data corresponding to an integrated luminosity of 3 $fb^{-1}$, the LHCb experiment gives the following upper limits at $95\%$ CL \cite{lhcb1},
\begin{eqnarray}
Br(B_d^0\rightarrow e\mu)<1.3\times 10^{-9},Br(B_s^0\rightarrow e\mu)<6.3\times 10^{-9},\nonumber
\end{eqnarray}
which substitutes for the previous results\cite{lhcb2}.

In this paper, we study the LFV decays of $B^0_{d,s}$ in a minimal extension of SM with NP featuring
extra scalar. The scalar is predicted by many extensions of SM and not observed yet even though many searches have been devoted to find it at the experiment. For simplicity, the couples of the neutral scalar with charged fermions are studied. We investigate the LFV decays of $B^0_{d,s}$ in a function of the couplings between the neutral scalar and quarks. We have considered the loop contributions for reason to
understand the different contributions from tree-level diagrams and
loop diagrams. It shows loop contributions is about two orders of
magnitude below tree-level contributions for $B^0_{d}$ and one order of
magnitude below tree-level contributions for $B^0_{s}$.

The paper is organized as follows. In Section \ref{sec2}, we provide a simple formalism for the description of the newly introduced scalar and give the analytic expression of LFV decays of $B^0_q\rightarrow\bar{l_l}l_k$ in detail. The numerical results are presented in Section \ref{sec3}, and the conclusion is drawn in Section \ref{sec4}.

\section{Formalism\label{sec2}}

In this section, we give the description of the minimal extension of the SM. In general, the minimal extension of the SM involves only one scalar, one vector or one fermion. In following we will consider the extension of the SM with one neutral singlet scalar $S^0$. We will add the neutral singlet scalar $S^0$ as a minimal extension of SM, that is, we are not concerned which models predict the new particle, but only want to investigate the observable phenomenon of this SM extension.

For simplicity and consideration of couplings like SM Higgs-fermion-fermion interactions, the couplings of `new' scalar and left-handed fermions are assumed equal
to that of the scalar and right-handed fermions. The interactions between the different generation up type quarks, down type quarks or charged leptons and the neutral scalar $S^0$ take the following structure,
\begin{eqnarray}
&&C_{u^iu^j}\bar{u^i}P_{L/R}u^j S^0,u^i\ne u^j,\nonumber\\
&&C_{d^id^j}\bar{d^i}P_{L/R}d^j S^0,d^i\ne d^j,\label{coeff}\\
&&C_{e^ie^j}\bar{e^i}P_{L/R}e^j S^0,e^i\ne e^j,\nonumber
\end{eqnarray}
and the couplings $C_{u^iu^j}$, $C_{d^id^j}$ and $C_{e^ie^j}$ are real numbers. The interactions between the same generation quarks or leptons are neglected and so the interactions between the gauge vectors or Higgs and the new scalar $S^0$ for simplicity. From Eq.(\ref{coeff}), one can see that the LFV decay originate from the interactions between different generation leptons and the new scalar $S^0$. The interactions between different generation quarks and the new scalar $S^0$ can contribute to the LFV decays $B^0_q\rightarrow\bar{l_l}l_k$ in quark sector.

\begin{figure}[htbp]
\setlength{\unitlength}{1mm}
\centering
\begin{minipage}[c]{1\textwidth}
\includegraphics[width=2.5in]{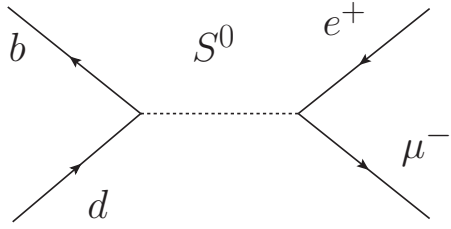}
\end{minipage}
\caption[]{Tree level diagrams contribute to $B^0_{d}\rightarrow e^+\mu^{-}$.}
\label{Tree}
\end{figure}

\begin{figure}[htbp]
\setlength{\unitlength}{1mm}
\centering
\begin{minipage}[c]{1\textwidth}
\includegraphics[width=5.0in]{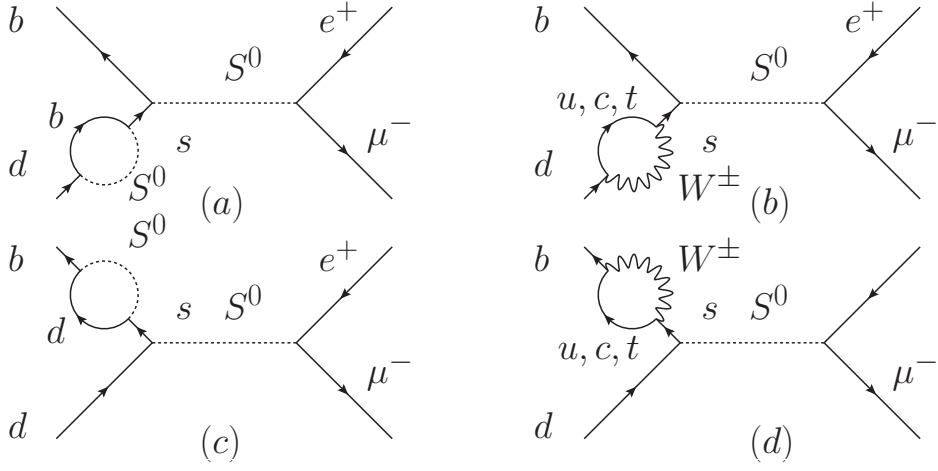}
\end{minipage}
\caption[]{Two point diagrams contribute to $B^0_{d}\rightarrow e^+\mu^{-}$.}
\label{twop}
\end{figure}

\begin{figure}[htbp]
\setlength{\unitlength}{1mm}
\centering
\begin{minipage}[c]{1.0\textwidth}
\includegraphics[width=5.3in]{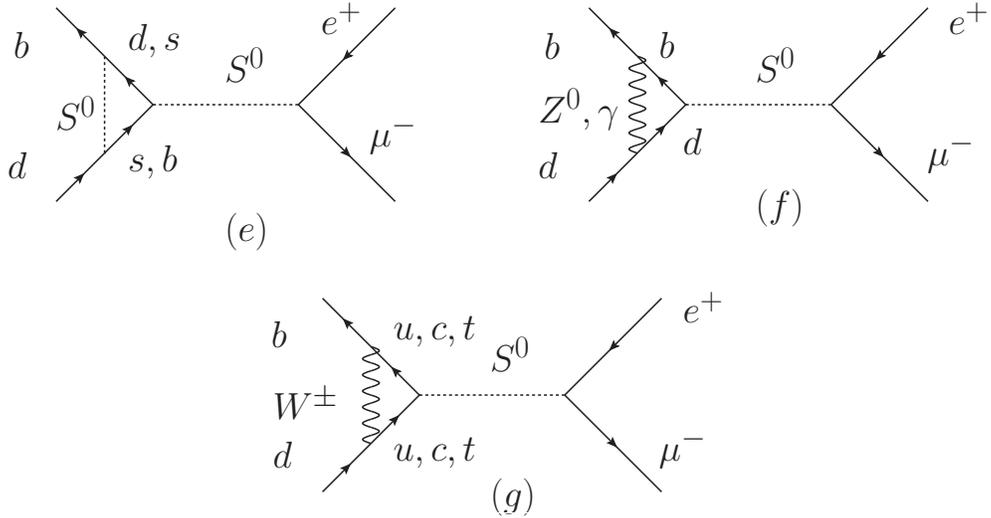}
\end{minipage}
\caption[]{Penguin diagrams contribute to $B^0_{d}\rightarrow e^+\mu^{-}$.}
\label{threep}
\end{figure}

\begin{figure}[htbp]
\setlength{\unitlength}{1mm}
\centering
\begin{minipage}[c]{1.0\textwidth}
\includegraphics[width=4.3in]{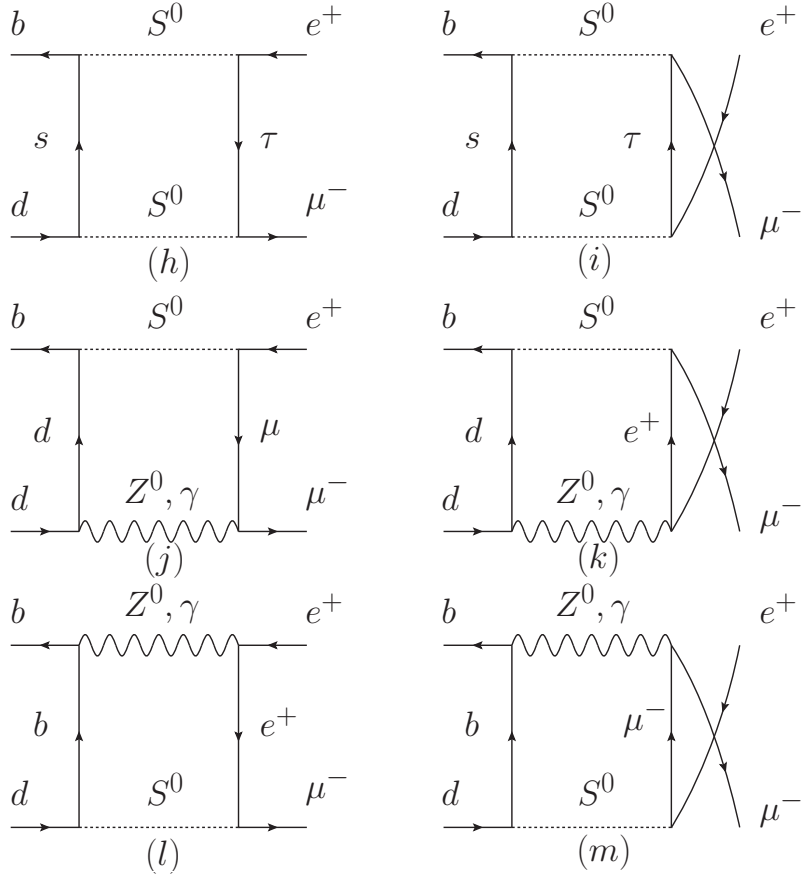}
\end{minipage}
\caption[]{Box diagrams contribute to $B^0_{d}\rightarrow e^+\mu^{-}$.}
\label{fourp}
\end{figure}

The one loop Feynman diagrams contributing to the LFV decay $B^0_{d}\rightarrow e^+\mu^{-}$ are presented in FIG.\ref{Tree}, FIG.\ref{twop}, FIG.\ref{threep} and FIG.\ref{fourp}, and other LFV decays of $B^0_{d}$ and $B^0_{s}$ can be discussed in a similar way.
Utilizing the notation of ref.\cite{Dedes}, the effective Hamiltonian of LFV decays of $B^0_{q}$($q = d, s$) is given by
\begin{eqnarray}
\mathcal{H}_{eff}=\frac{1}{16\pi^2}\sum_{X,Y=L,R}(C_{SXY}\mathcal{O}_{SXY}+C_{VXY}\mathcal{O}_{VXY}+C_{TX}\mathcal{O}_{TX}),
\end{eqnarray}
where $C_{SXY}$, $C_{VXY}$ and $C_{TX}$ are the Wilson coefficients. The relevant operators, including the scalar, vector and tensor operators, are given by,
\begin{eqnarray}
\mathcal{O}_{SXY}=(\bar{b}P_X q)(\bar{l_l}P_Y l_k),\mathcal{O}_{VXY}=(\bar{b}\gamma^{\mu}P_X q)(\bar{l_l}\gamma_{\mu}P_Y l_k),\mathcal{O}_{TX}=(\bar{b}\sigma^{\mu\nu}P_X q)(\bar{l_l}\sigma_{\mu\nu}l_k),
\label{SVXY}
\end{eqnarray}
where $l_l$ and $l_k$ are leptons, $P_{L/R}=\frac{1}{2}(1\mp\gamma_5)$. For $B^0_{d}\rightarrow e^+\mu^{-}$, symbols in Eq.(\ref{SVXY}) are $q$=$d$, $l_l$=$\mu$ and $l_k$=$e$. It is impossible to get a antisymmetric combination made up of $p^{\mu}$ by exchanging the index $\mu\leftrightarrow \nu$, so the tensor current $\left \langle 0|\bar{b}\sigma^{\mu\nu} q|B^0_{q}(p) \right \rangle$ vanishes. The expectation values of the matrix elements are derived as
\begin{eqnarray}
\left \langle 0|\bar{b}\gamma^\mu P_{L/R}q|B^0_{q}(p) \right \rangle=\mp\frac{i}{2}p^\mu f_{B^0_q},
\left \langle 0|\bar{b} P_{L/R}q|B^0_{q}(p) \right \rangle=\pm\frac{i{M^2_{B^0_q}} f_{B^0_q}}{2(m_b+m_q)},
\label{hardp}
\end{eqnarray}
where $f_{B^0_q}$ is the decay constant of $B^0_q$, $M_{B^0_q}$ is the mass of $B^0_q$.

Tree level diagrams contributing to $B^0_{d}\rightarrow e^+\mu^{-}$ is presented in FIG.\ref{Tree}. Using the equations in Eq.(\ref{hardp}), the relevant Wilson coefficient is calculated by,
\begin{eqnarray}
&&C_{SLL}=\frac{i C_{db}C_{e\mu}}{m^2_{S^0}-M^2_{B^0_{q}}},C_{SLR}=C_{SRL}=C_{SRR}=C_{SLL},\nonumber
\end{eqnarray}
where $m_{S^0}$ is the mass of neutral scalar $S^0$. One can see that only coefficient $C_{db}$ contributes to $B^0_{d} \rightarrow e^+\mu^{-}$ in quark sector.
%%%%%%%%%%%%%%%%%%%%%%%%%%%%%%%%%%%%%%%%%%%%%%%%%%%%%%%%%%%%%%%%%%%
%%%%%%%%%%%%%%%%%%%%%%%%%%%%%%%%%%%%%%%%%%%%%%%%%%%%%%%%%%%%%%%%%%%
Two point diagrams contributing to $B^0_{d}\rightarrow e^+\mu^{-}$ are presented in FIG.\ref{twop}. The relevant Wilson coefficient is calculated by,
%%%%%%%%%%%%%%%%%%%%%%%%%%%%%%%%%%%%%%%%%%%%%%%%%%%%%%%%%%%%%%%%%%%
%%%%%%%%%%%%%%%%%%%%%%%%%%%%%%%%%%%%%%%%%%%%%%%%%%%%%%%%%%%%%%%%%%%
%%%%%%%%%%%%%%%%%%%%%%%%%%%%%%%%%%%%%%%%%%%%%%%%%%%%%%%%%%%%%%%%%%%
%%%%%%%%%%%%%%%%%%%%%%%%%%%%%%%%%%%%%%%%%%%%%%%%%%%%%%%%%%%%%%%%%%%
%%%%%%%%%%%%%%%%%%%%%%%%%%%%%%%%%%%%%%%%%%%%%%%%%%%%%%%%%%%%%%%%%%%
\begin{eqnarray}
&&C^{(a)}_{SLL}=\frac{C_{db}C_{sb}^2 C_{e\mu}}{(m^2_{S^0}-M^2_{B^0_{q}})(m^2_s-m^2_d)}\big(
m_d(m_s-m_d)B_1(m_d^2;m_{S^0},m_b)+A_0(m_{S^0})\nonumber\\
&&\hspace{1.0cm}+(m_b^2+m_bm_s+m_dm_s+m_bm_d-m^2_{S^0})B_0(m_d^2;m_{S^0},m_b)-A_0(m_b)
\big),\nonumber\\
&&C^{(a)}_{SLR}=C^{(a)}_{SRL}=C^{(a)}_{SRR}=C^{(a)}_{SLL}.\nonumber
\end{eqnarray}
The Wilson coefficients for other two point diagrams are listed in APPENDIX.\ref{appa}. Coefficients $C_{ds}$,$C_{db}$ and $C_{sb}$ contribute to $B^0_{d} \rightarrow e^+\mu^{-}$ in quark sector. Penguin diagrams contributing to $B^0_{d}\rightarrow e^+\mu^{-}$ are presented in FIG.\ref{threep} and the corresponding Wilson coefficients are listed in APPENDIX.\ref{appb}. Coefficients $C_{ds}$, $C_{db}$, $C_{sb}$, $C_{uc}$,$C_{ut}$ and $C_{ct}$ contribute to the $B^0_{d} \rightarrow e^+\mu^{-}$ in quark sector at penguin diagram level. It is noted worthwhile that coefficients $C_{uc}$,$C_{ut}$ and $C_{ct}$ contribute to the LFV decays only at this level. Box diagrams contributing to $B^0_{d}\rightarrow e^+\mu^{-}$ are presented in FIG.\ref{fourp} and the corresponding Wilson coefficients are listed in APPENDIX.\ref{appc}. Coefficients $C_{ds}$,$C_{db}$ and $C_{sb}$ contribute to the $B^0_{d} \rightarrow e^+\mu^{-}$ in quark sector at box diagram level. All integrals in Wilson coefficients can be calculated by Package-X\cite{X}, which deals with analytic calculation and symbolic manipulation of one-loop Feynman integrals.

Every amplitude $\mathcal{M}$ in FIG.\ref{Tree},FIG.\ref{twop},FIG.\ref{threep} and FIG.\ref{fourp} is composed of the scalar, pseudoscalar, vector and axial-vector current, i.e.,
\begin{eqnarray}
(4\pi)^2\mathcal{M}=F_S\bar{l}_ll_k+F_P\bar{l}_l\gamma^5l_k+F_V p_{\mu}\bar{l}_l\gamma^\mu l_k+F_Ap_\mu\bar{l}_l\gamma^\mu\gamma^5 l_k,
\label{M}
\end{eqnarray}
where $l_l=\mu$ and $l_k=e$ for $B^0_{d} \rightarrow e^+\mu^{-}$ , and the form factors $F_S$, $F_P$, $F_V$ and $F_A$ are combinations of Wilson coefficients,
\begin{eqnarray}
F_S&=&\frac{i M^2_{B^0_q}f_{B^0_q}}{4(m_b+m_q)}(C_{SLL}+C_{SLR}-C_{SRR}-C_{SRL}),\nonumber\\
F_P&=&\frac{i M^2_{B^0_q}f_{B^0_q}}{4(m_b+m_q)}(-C_{SLL}+C_{SLR}-C_{SRR}+C_{SRL}),\nonumber\\
F_V&=&-\frac{i f_{B^0_q}}{4}(C_{VLL}+C_{VLR}-C_{VRR}-C_{VRL}),\nonumber\\
F_A&=&-\frac{i f_{B^0_q}}{4}(-C_{VLL}+C_{VLR}-C_{VRR}+C_{VRL}).\nonumber
\end{eqnarray}
From Eq.(\ref{M}) one can easily calculate the squared amplitude,
\begin{eqnarray}
|\mathcal{M}|^2&=&\frac{1}{128\pi^4}\big(|F_S|^2(M^2_{B^0_q}-(m_k+m_l)^2)+|F_P|^2(M^2_{B^0_q}-(m_l-m_k)^2)
\nonumber\\
&&+|F_V|^2(M^2_{B^0_q}(m_k-m_l)^2-(m_k-m_l)^2)+|F_A|^2(M^2_{B^0_q}(m_k+m_l)^2\nonumber\\
&&-(m_k-m_l)^2)+2 Re(F_S F_V^*)(m_l-m_k)(M^2_{B^0_q}+(m_k+m_l)^2)\nonumber\\
&&+2 Re(F_P F_A^*)(m_l+m_k)(M^2_{B^0_q}-(m_k-m_l)^2)\big).\nonumber
\end{eqnarray}
The analytic expression of the branching ratio of $B^0_q\rightarrow\bar{l_l}l_k$ is given by,
\begin{eqnarray}
Br(B^0_q\rightarrow\bar{l_l}l_k)=\frac{\tau_{B^0_q}}{16\pi M_{B^0_q}}\sqrt{1-(\frac{m_k+m_l}{M^2_{B^0_q}})^2}\sqrt{1-(\frac{m_k-m_l}{M^2_{B^0_q}})^2}
|\mathcal{M}|^2,
\end{eqnarray}
where $\tau_{B^0_q}$ is the life time of $B^0_q$.

\section{Numerical Analysis\label{sec3}}
\indent\indent

In the numerical analysis, we will adopt following values for parameters of meson $B^0_{d,s}$,
\begin{eqnarray}
&&m_{B^0_d}=5.279GeV,f_{B^0_d}=0.190GeV,\tau_{B^0_d}=1.520\times 10^{-12}s,\nonumber\\
&&m_{B^0_s}=5.366GeV,f_{B^0_s}=0.277GeV,\tau_{B^0_s}=1.509\times 10^{-12}s.\nonumber
\label{Bconstant}
\end{eqnarray}
The experimental bounds on LFV decays, such as radiative two body decays $l_i\rightarrow l_j\gamma$, leptonic three body decays $l_i\rightarrow 3l_j$ and $\mu-e$ conversion in nuclei, can give strong constraints on the coefficients $C_{e\mu}$, $C_{e\tau}$ and $C_{\mu\tau}$. In the following we will use LFV decays $l_i\rightarrow l_j\gamma$ to constrain the coefficients $C_{e\mu}$, $C_{e\tau}$ and $C_{\mu\tau}$.

\begin{figure}[htbp]
\setlength{\unitlength}{1mm}
\centering
\begin{minipage}[c]{1\textwidth}
\includegraphics[width=6.0in]{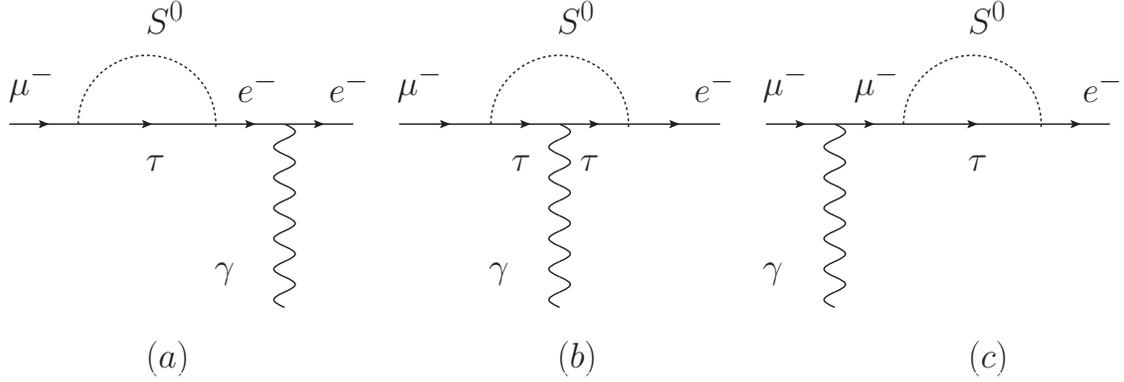}
\end{minipage}
\caption[]{The Feynman diagrams contributing to $\mu\rightarrow e\gamma$.}
\label{mutoe}
\end{figure}

The scalar $S^0$ mediated diagrams for $\mu\rightarrow e\gamma$ are shown in FIG.\ref{mutoe}. Taking account of the gauge invariance, and assuming the photon is on shell and transverse, the amplitude for $\mu\rightarrow e\gamma$ is given by\cite{DingY}
%%%%%%%%%%%%%%%%%%%%%%%%%%%%%%%%%%%%%%%%%%%%%%%%%%%%%%%%%%%%%%%%%%%
%%%%%%%%%%%%%%%%%%%%%%%%%%%%%%%%%%%%%%%%%%%%%%%%%%%%%%%%%%%%%%%%%%%
\begin{eqnarray}
M(\mu\rightarrow e\gamma)=\epsilon^{\mu\ast}\bar{u_e}(p_e)[iq^{\nu}\sigma_{\mu\nu}(A+B\gamma_5)]u_{\mu}(p_{\mu}).\nonumber
\end{eqnarray}
Then, the analytic expression of decay width is calculated by,
\begin{eqnarray}
\Gamma(\mu\rightarrow e \gamma)=\frac{(m_{\mu}^2-m_e^2)^3}{8\pi m_{\mu}^3}(|A|^2+|B|^2),\nonumber
\end{eqnarray}
where A is given by
\begin{eqnarray}
A&=&\frac{i e}{16\pi^2}C_{e\tau}C_{\mu\tau}\big( m_e (C_{22}(m^2_{\mu},0,m_e^2;m_{S^0},m_{\tau},m_{\tau})+C_{12}(m^2_{\mu},0,m_e^2;m_{S^0},m_{\tau},m_{\tau}))\nonumber\\
&&+m_{\mu} (C_{12}(m^2_{\mu},0,m_e^2;m_{S^0},m_{\tau},m_{\tau})+C_{11}(m^2_{\mu},0,m_e^2;m_{S^0},m_{\tau},m_{\tau}))+(m_e+m_{\tau})\nonumber\\
&&\times C_{2}(m^2_{\mu},0,m_e^2;m_{S^0},m_{\tau},m_{\tau})+(m_{\tau}+m_{\mu})C_{1}(m^2_{\mu},0,m_e^2;m_{S^0},m_{\tau},m_{\tau})\big).
\label{muAB}
\end{eqnarray}
and B equals zero. Actually, only FIG.\ref{mutoe}(b) contributes to the decay width cause the amplitudes in FIG.\ref{mutoe}(a) and FIG.\ref{mutoe}(c) are proportional to $\epsilon^{*\nu}\bar{u_e}(p_e)\gamma_{\nu}u_{\mu}(p_{\mu})$ or $\epsilon^{*\nu}\bar{u_e}(p_e)\gamma_{\nu}\gamma_5 u_{\mu}(p_{\mu})$. The decay width for $\tau\rightarrow e(\mu)\gamma$ can be formulated in a similar way. The integrals can also be calculated through the Package-X\cite{X}.

From Eq.(\ref{muAB}), one can see that the experimental bound of $\mu\rightarrow e\gamma$ can give constraint on coefficients $C_{e\tau}C_{\mu\tau}$. For $\tau\rightarrow e\gamma$ and $\tau\rightarrow \mu\gamma$, the coefficients $C_{e\mu}C_{\mu\tau}$ and $C_{e\mu}C_{e\tau}$ can be constrained.
Assuming the mass of the extra scalar $m_{S^0}$=13 TeV and taking account of the current limits of LFV decays $l_i\rightarrow l_j\gamma$ listed in TABLE \ref{tab2}, one can get the following values,
\begin{table}[h]
\caption{Current limits of LFV decays of $l_i\rightarrow l_j\gamma$. }
\begin{tabular}{@{}cccc@{}} \toprule
Decay&Bound&Decay&Bound\\
\colrule
$\mu\rightarrow e\gamma$&$4.2\times 10^{-13}$&$\tau\rightarrow e\gamma$&$3.3\times 10^{-8}$\\
$\tau\rightarrow \mu\gamma$&$4.4\times 10^{-8}$&-&-\\ \botrule
\end{tabular}
\label{tab2}
\end{table}
\begin{eqnarray}
C_{e\tau}C_{\mu\tau}\sim 4\times 10^{-5},C_{e\mu}C_{\mu\tau}\sim 6,C_{e\mu}C_{e\tau}\sim 60, \nonumber
\label{}
\end{eqnarray}
and easily calculate the result,
\begin{eqnarray}
C_{e\mu}\sim 3000,C_{e \tau}\sim 0.02,C_{\mu\tau}\sim 0.002.
\label{emt}
\end{eqnarray}
If not special specified, values in Eq.(\ref{emt}) are used as default in investigating the LFV decays of $B^0_{d,s}$.

%%%%%%%%%%%%%%%%%%%%%%%%%%%%%%%%%%%%%%%%%%%%%%%%%%%%%%%%%%%%%%%%%%%
\begin{figure}[htbp]
\setlength{\unitlength}{1mm}
\centering
\begin{minipage}[c]{1\columnwidth}
\includegraphics[width=1\columnwidth]{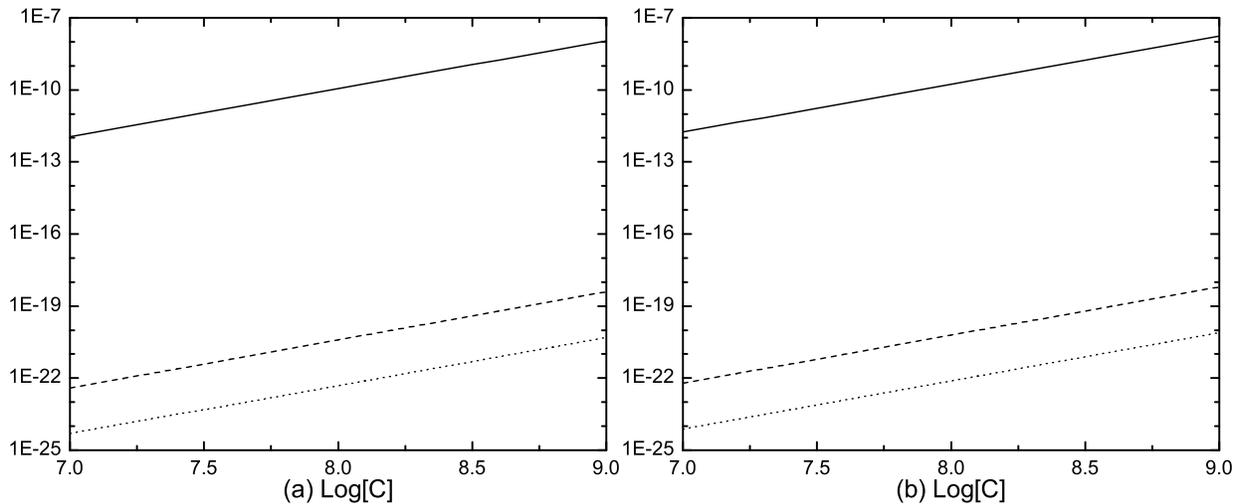}
\end{minipage}
\caption[]{(a)Br($B^0_{d}\rightarrow e^+\mu^-$)(solid line),Br($B^0_{d}\rightarrow e^+\tau^-$)(dash line) and Br($B^0_{d}\rightarrow \mu^+\tau^-$)(dot line) vs coefficient Log[C];(b)Br($B^0_{s}\rightarrow e^+\mu^-$) (solid line),Br ($B^0_{s}\rightarrow e^+\tau^-$)(dash line) and Br ($B^0_{s}\rightarrow \mu^+\tau^-$)(dot line) vs coefficient Log[C]. $C_{ds}$ = $C_{db}$ = $C_{sb}$ = $C_{uc}$ = $C_{ut}$ = $C_{ct}$ = C is assumed.}
\label{ds}
\end{figure}
%%%%%%%%%%%%%%%%%%%%%%%%%%%%%%%%%%%%%%%%%%%%%%%%%%%%%%%%%%%%%%%%%%%

In general case, we discuss the behavior of LFV decays of $B^0_{d,s}$ when all coefficients are universal. Taking $C_{ds}$ = $C_{db}$ = $C_{sb}$ = $C_{uc}$ = $C_{ut}$ = $C_{ct}$ = C, $m_{S^0}$ = 13 TeV, we plot the theoretical prediction of Br($B^0_{d}\rightarrow e^+\mu^-$) (solid line), Br ($B^0_{d}\rightarrow e^+\tau^-$)(dash line) and Br ($B^0_{d}\rightarrow \mu^+\tau^-$)(dot line) vs coefficient Log[C] in Fig.\ref{ds} (a) and the theoretical prediction of Br($B^0_{s}\rightarrow e^+\mu^-$) (solid line), Br ($B^0_{s}\rightarrow e^+\tau^-$)(dash line) and Br ($B^0_{s}\rightarrow \mu^+\tau^-$)(dot line) vs coefficient Log[C] in Fig.\ref{ds} (b). It shows that a linear relationship is displayed between LFV decays of $B^0_{d,s}$ and Log[C], and this displays the great dependence of LFV decays of $B^0_{d,s}$ on coefficient C. When coefficient C $\sim 10^9$, the prediction of Br($B^0_{d,s}\rightarrow e^+\mu^-$) is very close to the current limit in TABLE.\ref{tab1}. The prediction of LFV decays with outgoing $\tau$ lepton are far below the current limits.

\begin{figure}[htbp]
\setlength{\unitlength}{1mm}
\centering
\begin{minipage}[c]{1.0\columnwidth}
\includegraphics[width=1.0\columnwidth]{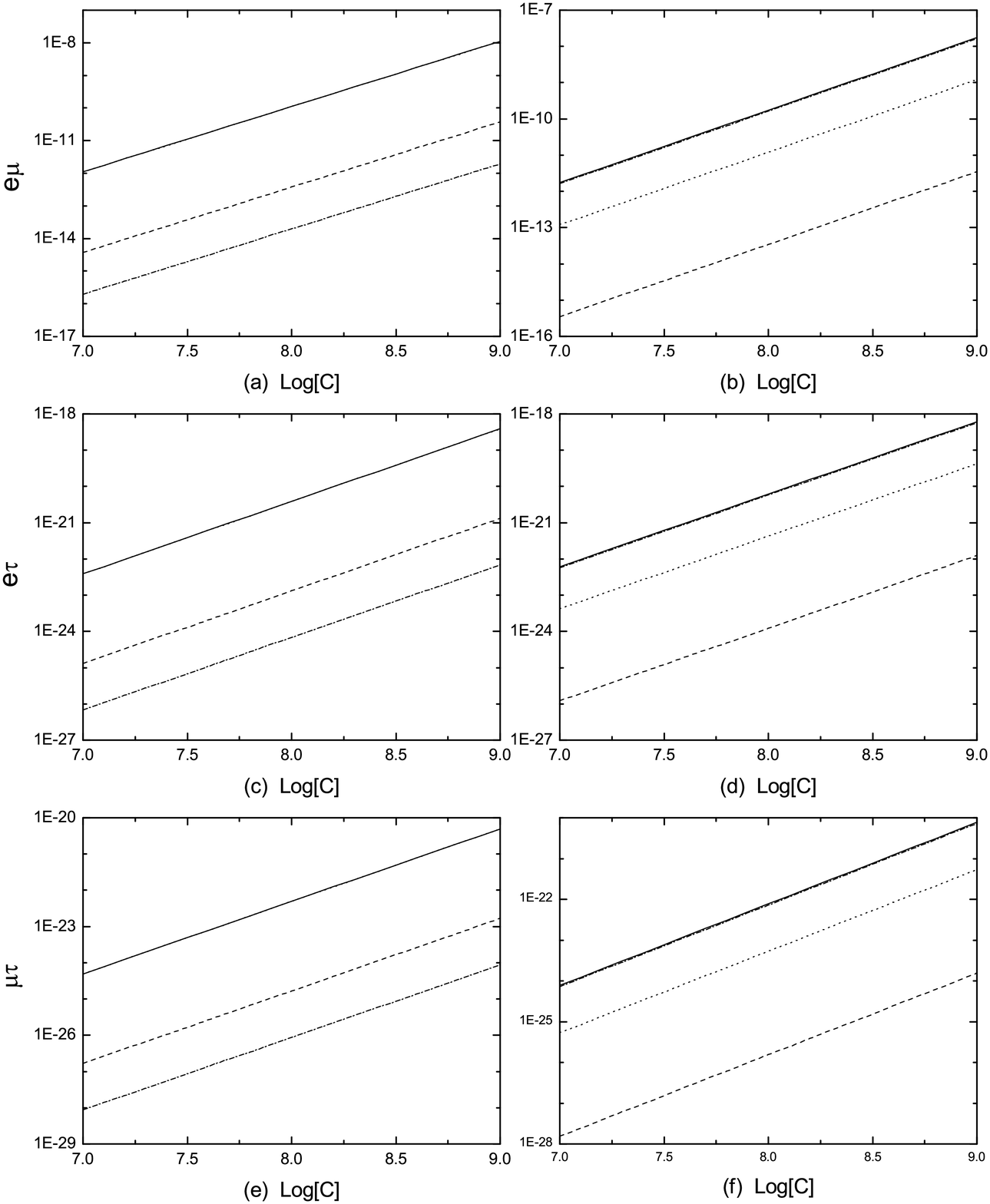}
\end{minipage}
\caption[]{Branching ratios of $B^0_{d,s}$ decay:(a)Br($B^0_{d}\rightarrow e^+\mu^-$),(b)Br($B^0_{d}\rightarrow e^+\tau^-$),(c)Br($B^0_{d}\rightarrow \mu^+\tau^-$),(d)Br($B^0_{s}\rightarrow e^+\mu^-$),(e)Br($B^0_{s}\rightarrow e^+\tau^-$) and (f)Br($B^0_{s}\rightarrow \mu^+\tau^-$). Following assumptions are used:(1)$C_{ds}$ = $C_{db}$ = $C_{sb}$ =C, $C_{uc}$ = $C_{ut}$ = $C_{ct}$ =0 (solid line),(2)$C_{ds}$ = C, $C_{db}$ = $C_{sb}$ =$C_{uc}$ = $C_{ut}$ = $C_{ct}$ =0 (dash line),(3)$C_{db}$ = C, $C_{ds}$ = $C_{sb}$ =$C_{uc}$ = $C_{ut}$ = $C_{ct}$ =0 (dot line),(4)$C_{sb}$ = C, $C_{ds}$ = $C_{db}$ =$C_{uc}$ = $C_{ut}$ = $C_{ct}$ =0 (dash dot line).}
\label{dsb}
\end{figure}

\begin{figure}[htbp]
\setlength{\unitlength}{1mm}
\centering
\begin{minipage}[c]{1.0\columnwidth}
\includegraphics[width=1\columnwidth]{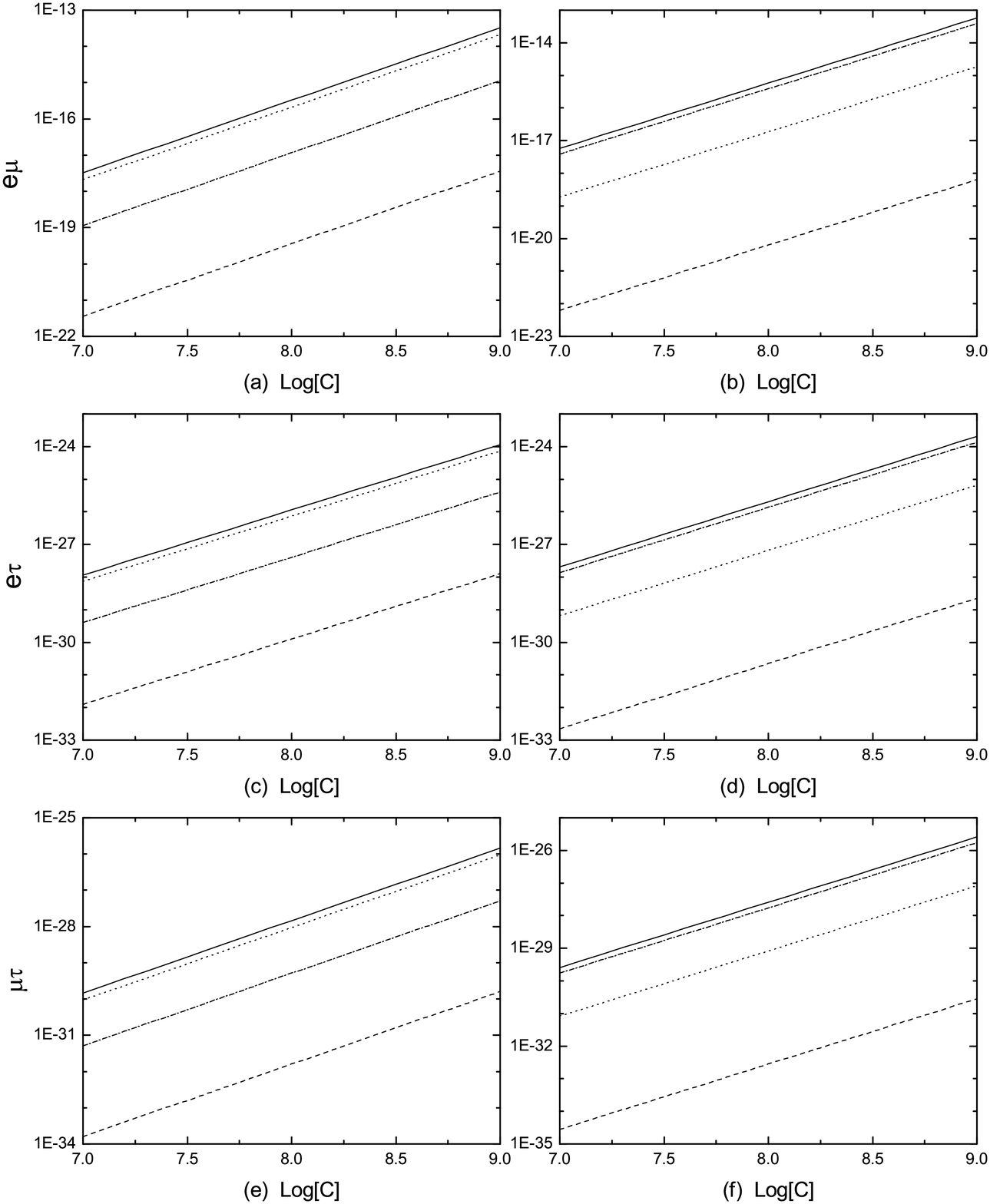}
\end{minipage}
\caption[]{Branching ratios of $B^0_{d,s}$ decay:(a)Br($B^0_{d}\rightarrow e^+\mu^-$),(b)Br($B^0_{d}\rightarrow e^+\tau^-$),(c)Br($B^0_{d}\rightarrow \mu^+\tau^-$),(d)Br($B^0_{s}\rightarrow e^+\mu^-$),(e)Br($B^0_{s}\rightarrow e^+\tau^-$) and (f)Br($B^0_{s}\rightarrow \mu^+\tau^-$). Following assumptions are used:(1)$C_{ds}$ = $C_{db}$ = $C_{sb}$ =0, $C_{uc}$ = $C_{ut}$ = $C_{ct}$ =C (solid line),(2)$C_{ds}$ = $C_{db}$ = $C_{sb}$ =0, $C_{uc}$ =C, $C_{ut}$ = $C_{ct}$ =0(dash line),(3)$C_{ds}$ = $C_{db}$ = $C_{sb}$ = $C_{uc}$ =0, $C_{ut}$ = C, $C_{ct}$ =0 (dot line),(4)$C_{ds}$ = $C_{db}$ = $C_{sb}$ = $C_{uc}$ = $C_{ut}$ = 0, $C_{ct}$ =C (dash dot line).}
\label{uct}
\end{figure}

Next, we investigate the LFV decays of $B^0_{d,s}$ in two cases: (I) Only the interactions between $S^0$ and down type quarks are considered, the interactions between $S^0$ and up type quarks are ignoring; (II) Only the interactions between $S^0$ and up type quarks are considered, the interactions between $S^0$ and down type quarks are ignoring. We also investigate the individual contributions from six coupling coefficients between quarks and new scalar, for example, by ignoring the solid
lines in FIG.\ref{dsb}(a) and FIG.\ref{uct}(a), then the rest three lines in FIG.\ref{dsb}(a) and three lines in FIG.\ref{uct}(a) are the six individual contributions from six coupling coefficients.

%%%%%%%%%%%%%%%%%%%%%%%%%%%%%%%%%%%%%%%%%%%%%%%%%%%%%%%%%%%%%%%%%%%
%%%%%%%%%%%%%%%%%%%%%%%%%%%%%%%%%%%%%%%%%%%%%%%%%%%%%%%%%%%%%%%%%%%

(I) In this case, coefficients $C_{uc}$,$C_{ut}$,$C_{ct}$ are set zero. Taking $C_{ds}$ = $C_{db}$ = $C_{sb}$ =C, we plot the theoretical prediction of LFV decays of $B^0_{d,s}$ vs coefficient Log[C] in Fig.\ref{dsb} (solid line). It shows the theoretical prediction of LFV decays of $B^0_{d,s}$ is very close to the general case. Then we plot the theoretical prediction of LFV decays of $B^0_{d,s}$ vs $C_{ds}$(dash line),$C_{db}$(dot line),$C_{sb}$(dash dot line) separately. It is manifest that coefficient $C_{db}$ dominates the LFV decays $B^0_{d}$ and so does coefficient $C_{sb}$ for the LFV decays $B^0_{s}$. Contributions from other coefficients are several orders of magnitudes below the condition in dominant coefficient. One can find reasons in FIG.\ref{Tree} that these LFV decays can appear in tree level where $C_{db}$ or $C_{sb}$ exist. The second coefficient dominates the LFV decays of $B^0_{d}$ is $C_{ds}$ and the coefficient $C_{sb}$ contributes the least among three coefficients. For the LFV decays of $B^0_{s}$, The second coefficient dominates the LFV decays is $C_{sb}$ and the last is $C_{ds}$.

%%%%%%%%%%%%%%%%%%%%%%%%%%%%%%%%%%%%%%%%%%%%%%%%%%%%%%%%%%%%%%%%%%%
%%%%%%%%%%%%%%%%%%%%%%%%%%%%%%%%%%%%%%%%%%%%%%%%%%%%%%%%%%%%%%%%%%%
(II) In this case, coefficients $C_{ds}$,$C_{db}$,$C_{sb}$ are set zero. Taking $C_{uc}$ = $C_{ut}$ = $C_{ct}$ =C, we plot the theoretical prediction of LFV decays of $B^0_{d,s}$ vs coefficient Log[C] in Fig.\ref{uct} (solid line). Different from case (I), it shows the theoretical prediction of LFV decays of $B^0_{d,s}$ is several orders of magnitude below the general case. Then we plot the theoretical prediction of LFV decays of $B^0_{d,s}$ vs $C_{uc}$(dash line),$C_{ut}$(dot line),$C_{ct}$(dash dot line) separately. It displays that the theoretical prediction for LFV decays of $B^0_{d}$ with three coefficients is $Br(C_{ut})>Br(C_{ct})>Br(C_{uc})$, and $Br(C_{ct})>Br(C_{ut})>Br(C_{uc})$ for LFV decays of $B^0_{s}$. This may be explained by the CKM matrix,
\begin{eqnarray}
|\mathcal{V}_{CKM}|=
\left(\begin{array}{ccc}
|V_{ud}|&|V_{us}|&|V_{ub}|\\
|V_{cd}|&|V_{cs}|&|V_{cb}|\\
|V_{td}|&|V_{ts}|&|V_{tb}|
\end{array}\right)=
\left(\begin{array}{ccc}
0.9742&0.2243&0.00394\\
0.218&0.997&0.0422\\
0.0081&0.0394&1.019
\end{array}\right).\nonumber
\label{CKM}
\end{eqnarray}
For meson $B^0_{d}$,
\begin{eqnarray}
&&\mathcal{M}(C_{uc})\propto |V_{ud}||V_{cb}|+|V_{ub}||V_{cd}|=0.0419702,\nonumber\\
&&\mathcal{M}(C_{ut})\propto |V_{ud}||V_{tb}|+|V_{ub}||V_{td}|=0.992742,\nonumber\\
&&\mathcal{M}(C_{ct})\propto |V_{cd}||V_{tb}|+|V_{cb}||V_{td}|=0.222484,\nonumber\\
&&\hspace{3.em}\mathcal{M}(C_{ut})>\mathcal{M}(C_{ct})>\mathcal{M}(C_{uc}).
\label{Mduct}
\end{eqnarray}
For meson $B^0_{s}$,
\begin{eqnarray}
&&\mathcal{M}(C_{uc})\propto |V_{us}||V_{cb}|+|V_{ub}||V_{cs}|=0.0133936,\nonumber\\
&&\mathcal{M}(C_{ut})\propto |V_{us}||V_{tb}|+|V_{ub}||V_{ts}|=0.228717,\nonumber\\
&&\mathcal{M}(C_{ct})\propto |V_{cs}||V_{tb}|+|V_{cb}||V_{ts}|=1.01761,\nonumber\\
&&\hspace{3.em}\mathcal{M}(C_{ct})>\mathcal{M}(C_{ut})>\mathcal{M}(C_{uc}).
\label{Msuct}
\end{eqnarray}
The orders listed in Eq.(\ref{Mduct}) and Eq.(\ref{Msuct}) coincide with the behavior displayed in FIG.\ref{uct}.

\section{Conclusions\label{sec4}}
In this work, taking account of the constraints on the parameter space from LFV decays Br($l_i\rightarrow l_j\gamma$), we analyze the LFV decays of $B^0_q\rightarrow\bar{l_l}l_k$ as a function of the six coefficients $C_{ds}$, $C_{db}$, $C_{sb}$, $C_{uc}$,$C_{ut}$ and $C_{ct}$ in the framework with one neutral single scalar introduced. The LFV decays of $B^0_d$ strongly depend on the magnitude of couplings between new scalar $S^0$ and the down type quarks, especially $C_{db}$ and the LFV decays of $B^0_s$ strongly depend on the magnitude of couplings $C_{sb}$. With one scalar $S^0$ introduced, the prediction on branching ratios of $B^0_{d,s}\rightarrow e^{\pm}\tau^{\mp}$ and $B^0_{d,s}\rightarrow \mu^{\pm}\tau^{\mp}$ are far below $B^0_{d,s}\rightarrow e^{\pm}\mu^{\mp}$ and the later are more promising to observed in future experiment.

\begin{acknowledgments}
\indent\indent
The work has been supported by the National Natural Science Foundation of China (NNSFC) with Grants No.11747064, Foundation of Department of Education of Liaoning province with Grant No. 2016TSPY10, Youth Foundation of the University of Science and Technology Liaoning with Grant No. 2016QN11.

\end{acknowledgments}
\appendix
\section{The Wilson coefficients for two point diagrams}
\label{appa}

\begin{eqnarray}
&&C^{(b)}_{SRR}=\sum_{i=u,c,t}\frac{e^2 C_{sb}C_{e\mu}K^{*}_{id} K_{is}m_d m_s}{2s_w^2m_{W}^2(m^2_{S^0}-M^2_{B^0_{q}})(m^2_s-m^2_d)}\big((m_d^2-m_i^2-2m_W^2)B_1(m_d^2;m_{W},m_i)\nonumber\\
&&\hspace{1.0cm}-m_W^2B_0(m_d^2;m_{W},m_i)+A_{0}(m_i)
\big),C^{(b)}_{SRL}=C^{(b)}_{SRR}\nonumber\\
&&C^{(b)}_{SLL}=\sum_{i=u,c,t}\frac{ e^2 C_{sb}C_{e\mu}K^{*}_{id} K_{is}}{2s_w^2m_W^2(m^2_{S^0}-M^2_{B^0_{q}})(m^2_s-m^2_d)}\big(
((m^2_i-m_d^2)^2+m^2_W(m_i^2-m_W^2))\nonumber\\
&&\hspace{1.0cm}\times B_0(m_d^2;m_{W},m_i)-m_d^2(m_W^2+m_i^2-m_d^2)B_1(m_d^2;m_{W},m_i)-(m_i^2+2m_W^2)\nonumber\\
&&\hspace{1.0cm}\times A_0(m_i)+(2m_W^2+m_i^2-m_d^2)A_0(m_W)\big),C^{(b)}_{SLR}=C^{(b)}_{SLL}\nonumber
\end{eqnarray}
\begin{eqnarray}
&&C^{(c)}_{SLL}=\frac{C_{db}C_{sd}^2 C_{e\mu}}{(m^2_{S^0}-M^2_{B^0_{q}})(m^2_b-m^2_s)}\big(
m_b(m_s+m_b)B_1(m_b^2;m_{S^0},m_d)-A_0(m_{S^0})\nonumber\\
&&\hspace{1.0cm}+(m^2_{S^0}-m_d^2+m_bm_s-m_dm_s+m_bm_d)B_0(m_b^2;m_{S^0},m_d)+A_0(m_d)
\big)\nonumber\\
&&C^{(c)}_{SLR}=C^{(c)}_{SRL}=C^{(c)}_{SRR}=C^{(c)}_{SLL}\nonumber
\end{eqnarray}
\begin{eqnarray}
&&C^{(d)}_{SLL}=\sum_{i=u,c,t}\frac{e^2 C_{sd}C_{e\mu}K_{ib} K^{*}_{is}m_bm_s}{2s_w^2m_{W}^2(m^2_{S^0}-M^2_{B^0_{q}})(m^2_b-m^2_s)}\big(m_W^2B_0(m_b^2;m_W,m_i)-A_0(m_i)\nonumber\\
&&\hspace{1.0cm}+(2m_W^2+m_i^2-m_b^2)B_{1}(m_b^2;m_W,m_i)
\big),C^{(d)}_{SLR}=C^{(d)}_{SLL}\nonumber\\
&&C^{(d)}_{SRR}=\sum_{i=u,c,t}\frac{e^2 C_{sd}C_{e\mu}K_{ib} K^{*}_{is}}{2s_w^2m_W^2(m^2_{S^0}-M^2_{B^0_{q}})(m^2_b-m^2_s)}\big((m_b^2-m_i^2-2m_W^2)A_0(m_W)
\nonumber\\
&&\hspace{1.0cm}-((m_b^2-m_i^2)^2+m_i^2m_W^2-2m_W^4)B_0(m_b^2;m_W,m_i)+(m_i^2+2m_W^2)A_0(m_i)\nonumber\\
&&\hspace{1.0cm}+m_b^2(2m_W^2+m_i^2-m_b^2)B_1(m_b^2;m_W,m_i)\big),C^{(d)}_{SRL}=C^{(d)}_{SRR}\nonumber
\end{eqnarray}
\section{The Wilson coefficients for penguin diagrams}
\label{appb}
\begin{eqnarray}
&&C^{(e)}_{SLL}=\sum_{i=d,s;j=s,b}^{i\ne j}\frac{C_{di}C_{jb}C_{ij}C_{e\mu}}{(m^2_{S^0}-M^2_{B^0_{q}})}\big(
B_0(M^2_{B^0_{q}};m_i,m_j)+(m_b+m_d-m_i+m_j)(m_dC_2\nonumber\\
&&\hspace{1.0cm}+m_bC_1)+(m_b(m_d+m_j)-m_dm_i-m_im_j+m^2_{S^0})C_0\big)\nonumber\\
&&C^{(e)}_{SLR}=C^{(e)}_{SRL}=C^{(e)}_{SRR}=C^{(e)}_{SLL},C_{\{0,1,2\}}=C_{\{0,1,2\}}(m_b^2,M^2_{B^0_{q}},m^2_d;m_{S^0},m_i,m_j)\nonumber
\end{eqnarray}

\begin{eqnarray}
&&C^{(f,Z)}_{SLL}=\frac{e^2C_{db}C_{e\mu}}{36c_w^2s_w^2m_Z^2(m^2_{S^0}-M^2_{B^0_{q}})}\big(
(4s_w^4-6s_w^2)A_0(m_Z)+4s_w^2m_Z^2(3-2s_w^2)B_0(m_b^2;m_Z,\nonumber\\
&&\hspace{1.0cm}m_b)+3m_bm_dB_0(M^2_{B^0_{q}};m_b,m_d)+4s_w^2m_Z^2(3-2s_w^2)B_0(m_d^2;m_Z,m_d)+6m_d^2s_w^2\nonumber\\
&&\hspace{1.0cm}\times B_1(m_b^2;m_Z,m_d)-m_b^2(8s_w^4-18s_w^2+9)B_1(m_b^2;m_Z,m_b)+m_Z^2(m_b^2(12s_w^2  \nonumber\\
&&\hspace{1.0cm}-8s_w^2)+m_bm_d(16s_w^4-36s_w^2+9)+4s_w^2(2s_w^2-3)(M^2_{B^0_{q}}-m_d^2))C^Z_0-2m_Z^2  \nonumber\\
&&\hspace{1.0cm}\times (m_b-m_d)(4m_ds_w^4C^Z_2+m_b(3-2s_w^2)^2C^Z_1)\big) \nonumber\\
&&C^{(f,Z)}_{SRL}=C^{(f)}_{SLL}+\frac{e^2C_{db}C_{e\mu}(4s_w^2-3)}{12c_w^2s_w^2m_Z^2(m^2_{S^0}-M^2_{B^0_{q}})}\big(
m_b^2B_0(m_b^2;m_Z,m_b)+2m_bm_dB_0(M^2_{B^0_{q}};m_Z,\nonumber\\
&&\hspace{1.0cm}m_d)+2m_Z^2m_b(m_b-m_d)C^Z_1+2m_Z^2m_d(m_d-m_b)C^Z_2-2m_bm_dm_Z^2C^Z_0\big)\nonumber\\
&&C^{(f,\gamma)}_{SLL}=\frac{2e^2C_{db}C_{e\mu}}{9M^2_{B^0_{q}}-m^2_{S^0}}\big(B_0(m_b^2;0,m_b)+B_0(m_d^2;0,m_d)
+(m_b-m_d)(m_bC^{\gamma}_1+m_dC^{\gamma}_{2})\nonumber\\
&&\hspace{1.0cm}+((m_b-m_d)^2-M^2_{B^0_{q}})C^{\gamma}_0\big)\nonumber\\
&&C^{(f,Z)}_{SLR}=C^{(f,Z)}_{SLL},C^{(f,Z)}_{SRR}=C^{(f,Z)}_{SRL},C^Z_{\{0,1,2\}}=C_{\{0,1,2\}}(m_b^2,M^2_{B^0_{q}},m_d^2;m_Z,m_b,m_d)\nonumber\\
&&C^{(f,\gamma)}_{SLR}=C^{(f,\gamma)}_{SLL}=C^{(f,\gamma)}_{SRL}=C^{(f,\gamma)}_{SRR},C^{\gamma}_{\{0,1,2\}}=C_{\{0,1,2\}}(m_b^2,M^2_{B^0_{q}},m_d^2;0,m_b,m_d)\nonumber
\end{eqnarray}
\begin{eqnarray}
&&C^{(g)}_{SLL}=\sum_{i,j=u,c,t}^{i\ne j}\frac{e^2C_{ij}C_{e\mu}K^{*}_{ib}K_{jd}m_b}{2s_w^2m_W^2(M^2_{B^0_{q}}-m^2_{S^0})}\big(
m_iB_1(m_b^2;m_W,m_i)+m_jB_0(M^2_{B^0_{q}};m_i,m_j)\nonumber\\
&&\hspace{1.0cm}+m_jm_W^2C_0+(m_b^2m_j-m_d^2m_i-(m_i-m_j)(m_im_j-2m_W^2))C_1\big)\nonumber\\
&&C^{(g)}_{SRR}=\sum_{i,j=u,c,t}^{i\ne j}\frac{e^2C_{ij}C_{e\mu}K^{*}_{ib}K_{jd}m_d}{2s_w^2m_W^2(m^2_{S^0}-M^2_{B^0_{q}})}\big(
m_iB_0(M^2_{B^0_{q}};m_i,m_j)+m_jB_1(m_b^2;m_W,m_i)\nonumber\\
&&\hspace{1.0cm}-m_im_W^2C_0-(m_b^2m_j-m_d^2m_i-(m_i-m_j)(m_im_j-2m_W^2))C_2\big)\nonumber\\
&&C^{(g)}_{SLR}=C^{(g)}_{SLL},C^{(g)}_{SRR}=C^{(g)}_{SRL},C_{\{0,1,2\}}=C_{\{0,1,2\}}(m_b^2,M^2_{B^0_{q}},m_d^2;m_W,m_i,m_j)\nonumber
\end{eqnarray}
\section{The Wilson coefficients for box diagrams}
\label{appc}
\begin{eqnarray}
&&C^{(h)}_{SLL}=C_{ds}C_{sb}C_{e\tau}C_{\mu\tau}(m_s(m_t+m_{\mu})D_0-m_d(m_{\tau}+m_{\mu})D_1-m_b(m_{\tau}+m_{\mu})D_2\nonumber\\
&&\hspace{1.0cm}-(m_d(m_{\tau}+m_{\mu})-m_sm_{\mu})D_3-m_dm_{\mu}D_{13}-m_bm_{\mu}D_{23}-m_dm_{\mu}D_{33})\nonumber\\
&&C^{(h)}_{VLL}=C_{ds}C_{sb}C_{e\tau}C_{\mu\tau}D_{00}\nonumber\\
&&C^{(h)}_{SLR}=C^{(h)}_{SLL}=C^{(h)}_{SRL}=C^{(h)}_{SRR},C^{(h)}_{VLR}=C^{(h)}_{VLL}=C^{(h)}_{VRL}=C^{(h)}_{VRR}\nonumber\\
&&D_{\{0,1,...\}}=D_{\{0,1,...\}}(m_d^2,M^2_{B^0_{q}},m_e^2,m_d^2+m_e^2-M^2_{B^0_{q}};m_b^2,m_{\mu}^2;m_s,m_{S^0},m_{S^0},m_{\tau})\nonumber
\end{eqnarray}
\begin{eqnarray}
&&C^{(i)}_{SLL}=C^{(i)}_{SLR}=C^{(i)}_{SRR}=C^{(i)}_{SRL}=C^{(h)}_{SLL}(m_e\leftrightarrow m_{\mu})\nonumber\\
&&C^{(i)}_{VLL}=C^{(i)}_{VLR}=C^{(i)}_{VRL}=C^{(i)}_{VRR}=C^{(h)}_{VLL}(m_e\leftrightarrow m_{\mu})\nonumber\\
&&D_{\{0,1,...\}}=D_{\{0,1,...\}}(m_d^2,M^2_{B^0_{q}},m_{\mu}^2,m_d^2+m_{\mu}^2-M^2_{B^0_{q}};m_b^2,m_{e}^2;m_s,m_{S^0},m_{S^0},m_{\tau})\nonumber
\end{eqnarray}
\begin{eqnarray}
&&C^{(j,Z)}_{SLL}=\frac{e^2C_{db}C_{e\mu}}{18c_w^2}\big((6-4s_w^2)C_0-2m_d^2(s_w^2-3)D_0+(M^2_{B^0_{q}}(4s_w^2-6)+3m_d^2-2(2s_w^2-3)\nonumber\\
&&\hspace{1.0cm}\times(m_e^2-m_{\mu}^2))D_3+(2s_w^2-3)(m_d^2D_{33}+2m_d^2D_{23}+(m_d^2+m_{\mu}^2)D_{22})+(M^2_{B^0_{q}}(4s_w^2-6)\nonumber\\
&&\hspace{1.0cm}+3m_d^2-2(2s_w^2-3)(m_e^2-m_{\mu}^2))D_2+(2s_w^2-3)(m_b^2D_{11}+(m_b^2-M^2_{B^0_{q}}+m_d^2)(D_{13}\nonumber\\
&&\hspace{1.0cm}+D_{12})+4D_{00})-3m_bm_dD_1\big),C^{(j,Z)}_{SLR}=C^{(j,Z)}_{SLL}\nonumber\\
&&C^{(j,Z)}_{SRR}=\frac{e^2C_{db}C_{e\mu}}{18c_w^2}\big(2m_d^2s_w^2(D_{33}+2D_{23})-4s_w^2C_0-m_d^2(2s_w^2-3)D_0+(4s_w^2(M^2_{B^0_{q}}-m_e^2\nonumber\\
&&\hspace{1.0cm}+m_{\mu}^2)-3m_d^2)D_3+4M^2_{B^0_{q}}s_w^2D_2-4m_e^2s_w^2D_2-3m_d^2D_2+2s_w^2((m_d^2+m_{\mu}^2)D_{22}+m_b^2D_{11}\nonumber\\
&&\hspace{1.0cm}+(m_b^2-M^2_{B^0_{q}}+m_d^2)(D_{13}+D_{12})+4D_{00})+3m_bm_dD_1\big),C^{(j,Z)}_{SRL}=C^{(j,Z)}_{SRR}\nonumber\\
&&C^{(j,Z)}_{VLL}=\frac{e^2C_{db}C_{e\mu}m_{\mu}}{18c_w^2}\big(2m_ds_w^2(D_{23}+D_{22})+(3m_b-2m_bs_w^2)D_{12}+m_d(2s_w^2-3)D_2\big)\nonumber\\
&&C^{(j,Z)}_{VRR}=\frac{e^2C_{db}C_{e\mu}m_{\mu}}{18c_w^2}\big(2m_ds_w^2(D_{2}+D_{22})-2m_bs_w^2D_{12}+m_d(2s_w^2-3)D_{23}-3m_dD_{22}\big)\nonumber\\
&&C^{(j,Z)}_{VLR}=C^{(j,Z)}_{VLL},C^{(j,Z)}_{VRL}=C^{(j,Z)}_{VRR},C_{\{0\}}=C_{\{0\}}(m_e^2,m_{\mu}^2,M^2_{B^0_{q}};m_{S^0},m_{\mu},m_Z)\nonumber\\
&&D_{\{0,1,...\}}=D_{\{0,1,...\}}(m_b^2,m_e^2,m_{\mu}^2,m_d^2;m_d^2+m_e^2-M^2_{B^0_{q}},M^2_{B^0_{q}};m_d,m_{S^0},m_{\mu},m_Z)\nonumber
\end{eqnarray}

\begin{eqnarray}
&&C^{(j,\gamma)}_{SLL}=\frac{e^2C_{db}C_{e\mu}}{3}\big(m_d^2(D_{33}+2D_{23}-D_0)+(m_d^2+m_{\mu}^2)D_{22}+2(M^2_{B^0_{q}}-m_{e}^2)D_2-2C_0\nonumber\\
&&\hspace{1.0cm}+2(M^2_{B^0_{q}}+m_{\mu}^2-m_{e}^2)D_3+m_b^2D_{11}+(m_{b}^2+m_d^2-M^2_{B^0_{q}})(D_{13}+D_{12})+4D_{00}\big)\nonumber\\
&&C^{(j,\gamma)}_{SLL}=\frac{e^2C_{db}C_{e\mu}m_{\mu}}{3}\big(m_dD_{23}+m_dD_{22}-m_bD_{12}+m_dD_{2}\big)\nonumber\\
&&C^{(j,\gamma)}_{SLR}=C^{(j,\gamma)}_{SLL}=C^{(j,\gamma)}_{SRL}=C^{(j,\gamma)}_{SRR},C^{(j,\gamma)}_{VLR}=C^{(j,\gamma)}_{VLL}=C^{(j,\gamma)}_{VRL}=C^{(j,\gamma)}_{VRR}\nonumber\\
&&C_0=C_0(m_e^2,m_{\mu}^2,M^2_{B^0_{q}};m_{S^0},m_{\mu},0)\nonumber\\
&&D_{\{0,1,...,33\}}=D_{\{0,1,...,33\}}(m_b^2,m_{e}^2,m_{\mu}^2,m_d^2;m_{e}^2+m_d^2-M^2_{B^0_{q}},M^2_{B^0_{q}};m_d,m_{S^0},m_{\mu},0)\nonumber
\end{eqnarray}
\begin{eqnarray}
&&C^{(k,Z)}_{SLL}=\frac{e^2C_{db}C_{e\mu}}{6c_w^2}\big((2s_w^2-3)(m_e^2D_{33}+2(M^2_{B^0_{q}}+m_e^2-m_{\mu}^2)D_0+2(M^2_{B^0_{q}}+2m_e^2-m_{\mu}^2)D_3)\nonumber\\
&&\hspace{1.0cm}+(2s_w^2(3M^2_{B^0_{q}}-m_b^2+m_d^2+2m_e^2-2m_{\mu}^2)+3m_b(m_b+m_d)-9M^2_{B^0_{q}}-6m_e^2\nonumber\\
&&\hspace{1.0cm}+6m_{\mu}^2)D_2+M^2_{B^0_{q}}(2s_w^2-3)D_{22}+2(2s_w^2-3)(M^2_{B^0_{q}}+m_e^2-m_{\mu}^2)D_{23}+(M^2_{B^0_{q}}(4s_w^2-6)\nonumber\\
&&\hspace{1.0cm}+m_d^2(4s_w^2-3)+2(2s_w^2-3)(m_e^2-m_{\mu}^2))D_1+(2s_w^2-3)(m_d^2D_{11}+2(M^2_{B^0_{q}}+m_e^2\nonumber\\
&&\hspace{1.0cm}-m_{\mu}^2)D_{13}+(M^2_{B^0_{q}}+m_d^2-m_{b}^2)D_{12}+4D_{00})\big),C^{(k,Z)}_{SLR}=C^{(k,Z)}_{SLL}\nonumber\\
&&C^{(k,Z)}_{SRR}=\frac{e^2C_{db}C_{e\mu}}{6c_w^2}\big(s_w^2(2m_e^2D_{33}+(M^2_{B^0_{q}}+m_e^2-m_{\mu}^2)D_0+4(M^2_{B^0_{q}}+2m_e^2-m_{\mu}^2)D_3)+(2s_w^2\nonumber\\
&&\hspace{1.0cm}\times(3M^2_{B^0_{q}}-m_b^2+m_d^2+2m_e^2-2m_{\mu}^2)-3m_d(m_b+m_d))D_2+4s_w^2(M^2_{B^0_{q}}+m_e^2-m_{\mu}^2)\nonumber\\
&&\hspace{1.0cm}D_{23}+2M^2_{B^0_{q}}s_w^2D_{22}+(4s_w^2(M^2_{B^0_{q}}+m_e^2-m_{\mu}^2)+m_d^2(4s_w^2-3))D_1+2s_w^2(m_d^2D_{11}+2\nonumber\\
&&\hspace{1.0cm}\times (M^2_{B^0_{q}}+m_e^2-m_{\mu}^2)D_{13}+(M^2_{B^0_{q}}+m_d^2-m_{b}^2)D_{12}+4D_{00})\big),C^{(k,Z)}_{SRL}=C^{(k,Z)}_{SRR}\nonumber\\
&&C^{(k,Z)}_{VLL}=C^{(k,Z)}_{VLR}=\frac{e^2C_{db}C_{e\mu}m_e}{6c_w^2}\big((3m_b-2m_bs_w^2-2m_ds_w^2)D_{23}-2m_ds_w^2D_{13}-3m_dD_{3}\big)\nonumber\\
&&C^{(k,Z)}_{VRR}=C^{(k,Z)}_{VRL}=\frac{e^2C_{db}C_{e\mu}m_e}{6c_w^2}\big((m_d(3-2s_w^2)-2m_bs_w^2)D_{23}+m_d((3-2s_w^2)D_{13}+3D_3)\big)\nonumber\\
&&D_{\{0,1,...\}}=D_{\{0,1,...\}}(m_d^2,m_b^2,m_{\mu}^2,m_e^2;M^2_{B^0_{q}},m_d^2+m_{\mu}^2-M^2_{B^0_{q}};m_Z,m_d,m_{S^0},m_e)\nonumber
\end{eqnarray}
\begin{eqnarray}
&&C^{(k,\gamma)}_{SLL}=\frac{e^2C_{db}C_{e\mu}}{3}\big(m_d^2(D_{33}-D_0)-(M^2_{B^0_{q}}-2m_d^2+m_e^2-m_{\mu}^2)D_{23}+4D_{00}+m_b^2D_{11}\nonumber\\
&&\hspace{1.0cm}-2(M^2_{B^0_{q}}+m_e^2-m_{\mu}^2)D_3-(M^2_{B^0_{q}}-m_e^2+m_{\mu}^2)D_2-(M^2_{B^0_{q}}-m_d^2-m_{\mu}^2)D_{22}\nonumber\\
&&\hspace{1.0cm}-(M^2_{B^0_{q}}-m_d^2-m_{b}^2)(D_{13}+D_{12})\big)\nonumber\\
&&C^{(k,\gamma)}_{VLL}=\frac{e^2C_{db}C_{e\mu}m_e}{3}\big(m_dD_{23}+m_dD_{22}-m_bD_{12}+m_dD_{2}\big)\nonumber\\
&&C^{(k,\gamma)}_{SLR}=C^{(k,\gamma)}_{SLL}=C^{(k,\gamma)}_{SRL}=C^{(k,\gamma)}_{SRR},C^{(k,\gamma)}_{VLR}=C^{(k,\gamma)}_{VLL}=C^{(k,\gamma)}_{VRL}=C^{(k,\gamma)}_{VRR}\nonumber\\
&&D_{\{0,...\}}=D_{\{0,...\}}(m_b^2,m_{\mu}^2,m_e^2,m_d^2;m_{\mu}^2+m_d^2-M^2_{B^0_{q}},M^2_{B^0_{q}};m_d,m_{S^0},m_{e},0)\nonumber
\end{eqnarray}
\begin{eqnarray}
&&C^{(l,Z)}_{SLL}=\frac{-e^2C_{db}C_{e\mu}}{6c_w^2}\big(2s_w^2((m_d^2+m_e^2-M^2_{B^0_{q}})D_{33}-m_bm_dD_0+(m_d(m_d-m_b)-M^2_{B^0_{q}}+m_e\nonumber\\
&&\hspace{1.0cm}\times (m_e+m_{\mu}))D_3)+(3m_b(m_b+m_d)-2s_w^2(m_d(m_b-m_d)+M^2_{B^0_{q}}))D_2+2s_w^2(m_b^2D_{22}\nonumber\\
&&\hspace{1.0cm}+(m_b^2+m_d^2-M^2_{B^0_{q}})D_{23}-(M^2_{B^0_{q}}-2m_d^2-m_e^2+m_{\mu}^2)D_{13}+m_d(m_d-m_b)D_1+4D_{00}\nonumber\\
&&\hspace{1.0cm}\times m_d^2D_{11}+(m_b^2+m_d^2-M^2_{B^0_{q}})D_{12})\big),C^{(l,Z)}_{SLR}=C^{(l,Z)}_{SLL}\nonumber\\
&&C^{(l,Z)}_{SRR}=\frac{e^2C_{db}C_{e\mu}}{6c_w^2}\big((2s_w^2-3)((M^2_{B^0_{q}}-m_d^2-m_e^2)D_{33}+m_bm_dD_0+(m_d(m_b-m_d)+M^2_{B^0_{q}}\nonumber\\
&&\hspace{1.0cm}-m_e(m_e+m_{\mu}))D_3)+(3m_b^2+2m_bm_ds_w^2+(2s_w^2-3)(M^2_{B^0_{q}}-m_d^2))D_2-(2s_w^2-3)\nonumber\\
&&\hspace{1.0cm}\times(m_b^2D_{22}+(m_b^2-M^2_{B^0_{q}}+m_d^2)D_{23}+m_d(m_d-m_b)D_1)+(2s_w^2-3)(-m_d^2D_{11}\nonumber\\
&&\hspace{1.0cm}+(M^2_{B^0_{q}}-2m_d^2-m_e^2+m_{\mu}^2)D_{13}-(m_b^2-M^2_{B^0_{q}}+m_d^2)D_{12}-4D_{00})\big),C^{(l,Z)}_{SRL}=C^{(l,Z)}_{SRR}\nonumber\\
&&C^{(l,Z)}_{VLL}=\frac{-e^2C_{db}C_{e\mu}}{6c_w^2}\big((2s_w^2-3)(-m_dm_{\mu}D_{33}-(m_d(m_e+m_{\mu})-m_bm_{\mu})D_3)+2s_w^2m_b(m_e\nonumber\\
&&\hspace{1.0cm}+m_{\mu})D_2+m_b(2s_w^2-3)(m_e+m_{\mu})D_0+2m_bm_{\mu}s_w^2D_{23}-m_d(2s_w^2-3)(m_{\mu}D_{13}\nonumber\\
&&\hspace{1.0cm}+(m_e+m_{\mu})D_1)\big),C^{(l,Z)}_{VLR}=C^{(l,Z)}_{VLL},C^{(l,Z)}_{VRL}=C^{(l,Z)}_{VRR}\nonumber\\
&&C^{(l,Z)}_{VRR}=\frac{e^2C_{db}C_{e\mu}}{6c_w^2}\big(2s_w^2(m_dm_{\mu}D_{33}-m_b(m_e+m_{\mu})D_0+(m_d(m_e+m_{\mu})-m_bm_{\mu})D_3)\nonumber\\ &&\hspace{1.0cm}+m_b(3-2s_w^2)((m_e+m_{\mu})D_2+m_{\mu}D_{23})+2m_ds_w^2(m_{\mu}D_{13}+(m_e+m_{\mu})D_1)\big)\nonumber\\
&&D_{\{0,1,...\}}=D_{\{0,1,...\}}(m_d^2,M^2_{B^0_{q}},m_e^2, m_d^2+m_{e}^2-M^2_{B^0_{q}};m_b^2,m_{\mu}^2;m_b,m_{S^0},m_Z,m_e)\nonumber
\end{eqnarray}
\begin{eqnarray}
&&C^{(l,\gamma)}_{SLL}=\frac{e^2C_{db}C_{e\mu}}{3}\big((m_d^2+m_{\mu}^2)D_{33}-2C_0+(m_d(m_b+2m_d)-2m_{S^0}^2)D_0+m_b^2D_{22}\nonumber\\
&&\hspace{1.0cm}+m_d^2D_{11}+(m_d(m_b+3m_d)-m_{\mu}(m_e+m_{\mu}))D_3+(2m_b^2+m_bm_d-M^2_{B^0_{q}}+m_d^2)\nonumber\\
&&\hspace{1.0cm}\times D_2+4D_{00}+(m_b^2+m_d^2-M^2_{B^0_{q}})(D_{23}+D_{12})+m_d(2m_dD_{13}+(m_b+3m_d)D_1)\big)\nonumber\\
&&C^{(l,\gamma)}_{VLL}=\frac{e^2C_{db}C_{e\mu}}{3}\big(m_dm_{\mu}D_{33}-m_b(m_e+m_{\mu})D_0+(m_d(m_e+m_{\mu})-m_bm_{\mu})D_3\nonumber\\
&&\hspace{1.0cm}-m_bm_{\mu}D_{23}-m_b(m_e+m_{\mu})D_2+m_dm_{\mu}D_{13}+m_d(m_e+m_{\mu})D_1\big)\nonumber\\
&&C^{(l,\gamma)}_{SLR}=C^{(l,\gamma)}_{SLL}=C^{(l,\gamma)}_{SRL}=C^{(l,\gamma)}_{SRR},C^{(l,\gamma)}_{VLR}=C^{(l,\gamma)}_{VLL}=C^{(l,\gamma)}_{VRL}=C^{(l,\gamma)}_{VRR}\nonumber\\
&&C_0=C_0(m_b^2,m_{e}^2,m_d^2+m_e^2-M^2_{B^0_{q}};m_{b},0,m_e)\nonumber\\
&&D_{\{0,...\}}=D_{\{0,...\}}(m_d^2,m_{M^2_{B^0_{q}}}^2,m_e^2,m_{e}^2+m_d^2-M^2_{B^0_{q}};m_b^2,m_{\mu}^2;m_b^2,m_{S^0},0,m_{e})\nonumber
\end{eqnarray}
\begin{eqnarray}
&&C^{(m,Z)}_{SLL}=\frac{e^2C_{db}C_{e\mu}}{6c_w^2}\big(2s_w^2((m_d^2+m_{\mu}^2-M^2_{B^0_{q}})D_{33}-m_bm_dD_{0}+(m_d^2-m_dm_b-M^2_{B^0_{q}}+m_{\mu}\nonumber\\
&&\hspace{1.0cm}\times(m_e+m_{\mu}))D_{3})+(3m_b(m_b+m_d)-2s_w^2(m_d(m_b-m_d)+M^2_{B^0_{q}}))D_2+2s_w^2(m_b^2\nonumber\\
&&\hspace{1.0cm}\times D_{22}+(m_b^2+m_d^2-M^2_{B^0_{q}})D_{23}-(M^2_{B^0_{q}}-2m_d^2+m_e^2-m_{\mu}^2)D_{13}+m_d(m_d-m_b)D_1)\nonumber\\
&&\hspace{1.0cm}+2s_w^2(m_d^2D_{11}+(m_b^2+m_d^2-M^2_{B^0_{q}})D_{12}+4D_{00})\big),C^{(m,Z)}_{SLR}=C^{(m,Z)}_{SLL}\nonumber\\
&&C^{(m,Z)}_{SRR}=\frac{e^2C_{db}C_{e\mu}}{6c_w^2}\big((2s_w^2-3)((m_d^2+m_{\mu}^2-M^2_{B^0_{q}})D_{23}-m_bm_dD_0+(m_d^2-m_dm_b-M^2_{B^0_{q}}\nonumber\\
&&\hspace{1.0cm}+m_{\mu}(m_e+m_{\mu}))D_3+m_b^2D_{22}+(2m_d^2-M^2_{B^0_{q}}-m_e^2+m_{\mu}^2)D_{13}+(m_b^2+m_d^2-M^2_{B^0_{q}})\nonumber\\
&&\hspace{1.0cm}\times (D_{12}+D_{13})+m_d^2D_{11}+4D_{00}+m_d(m_d-m_b)D_1)-(3m_b^2+2m_bm_ds_w^2\nonumber\\
&&\hspace{1.0cm}+(2s_w^2-3)(M^2_{B^0_{q}}-m_d^2))D_2\big),C^{(m,Z)}_{SRL}=C^{(m,Z)}_{SRR}\nonumber\\
&&C^{(m,Z)}_{VLL}=\frac{-e^2C_{db}C_{e\mu}}{6c_w^2}\big((3-2s_w^2)(m_dm_eD_{33}-m_b(m_e+m_{\mu})D_0+(m_bm_e-m_d(m_e+m_{\mu}))\nonumber\\
&&\hspace{1.0cm}\times D_3-m_dm_eD_{13}-m_d(m_e+m_{\mu})D_1))+2m_bm_es_w^2D_{23}+2m_bs_w^2(m_e+m_{\mu})D_2\big)\nonumber\\
&&C^{(m,Z)}_{VRR}=\frac{e^2C_{db}C_{e\mu}}{6c_w^2}\big(2s_w^2(m_d(m_eD_{33}+m_eD_{13}+(m_e+m_{\mu})D_1)-m_b(m_e+m_{\mu})D_0\nonumber\\ &&\hspace{1.0cm}+(m_d(m_e+m_{\mu})-m_bm_e)D_3)-m_b(2s_w^2-3)(m_eD_{23}+(m_e+m_{\mu})D_2)\big)\nonumber\\
&&C^{(m,Z)}_{VLR}=C^{(m,Z)}_{VLL},C^{(m,Z)}_{VRL}=C^{(m,Z)}_{VRR}\nonumber\\
&&D_{\{0,1,...\}}=D_{\{0,1,...\}}(m_d^2,M^2_{B^0_{q}},m_{\mu}^2, m_d^2+m_{\mu}^2-M^2_{B^0_{q}};m_b^2,m_{e}^2;m_b,m_{S^0},m_Z,m_{\mu})\nonumber
\end{eqnarray}
\begin{eqnarray}
&&C^{(m,\gamma)}_{SLL}=\frac{e^2C_{db}C_{e\mu}}{3}\big((m_d^2+m_{\mu}^2-M^2_{B^0_{q}})D_{33}-m_bm_dD_0-(m_d(m_b-m_d)+M^2_{B^0_{q}})D_2+m_b^2D_{22}\nonumber\\
&&\hspace{1.0cm}+(m_d(m_d-m_b)-M^2_{B^0_{q}}+m_{\mu}(m_e+m_{\mu}))D_3+(m_b^2-M^2_{B^0_{q}}+m_d^2)D_{23}+m_d(m_d\nonumber\\
&&\hspace{1.0cm}-m_b)D_1+m_d^2D_{11}-(M^2_{B^0_{q}}-2m_d^2+m_e^2-m_{\mu}^2)D_{13}+(m_b^2-M^2_{B^0_{q}}+m_d^2)D_{12}+4D_{00}
\big)\nonumber\\
&&C^{(m,\gamma)}_{VLL}=\frac{e^2C_{db}C_{e\mu}}{3}\big(m_dm_eD_{33}-m_b(m_e+m_{\mu})D_0+(m_d(m_e+m_{\mu})-m_bm_e)D_3-m_b\nonumber\\
&&\hspace{1.0cm}\times (m_e+m_{\mu})D_2-m_bm_eD_{23}+m_dm_eD_{13}+m_d(m_e+m_{\mu})D_1\big)\nonumber\\
&&C^{(m,\gamma)}_{SLR}=C^{(m,\gamma)}_{SLL}=C^{(m,\gamma)}_{SRL}=C^{(m,\gamma)}_{SRR},C^{(m,\gamma)}_{VLR}=C^{(m,\gamma)}_{VLL}=C^{(m,\gamma)}_{VRL}=C^{(m,\gamma)}_{VRR}\nonumber\\
&&D_{\{0,...\}}=D_{\{0,...\}}(m_d^2,m_{M^2_{B^0_{q}}}^2,m_{\mu}^2,m_{\mu}^2+m_d^2-M^2_{B^0_{q}};m_b^2,m_{e}^2;m_b^2,m_{S^0},0,m_{\mu})\nonumber
\end{eqnarray}


\begin{thebibliography}{99}



\bibitem{GUT1}
J.C. Pati and A. Salam, Phys. Rev. D10(1974)275.
\bibitem{GUT2}
H. Georgi and S.L. Glashow, Phys. Rev. Lett.32(1974)438.
\bibitem{GUT3}
P. Langacker, Phys. Rep.72(1981)185.

\bibitem{SUSY1}
H.E. Haber and G.L. Kane, Phys. Rep.117(1985)75.
\bibitem{SUSY2}
C.-H. Chang, T.-F. Feng, Eur. Phys. J.C12(2000)137.


\bibitem{sterile1}
A. Ilakovac, Phys. Rev. D62(2000)036010.
\bibitem{sterile2}
J. I. Illana and T. Riemann, Phys. Rev. D63(2001)053004.
\bibitem{sterile3}
A. Abada, V. De Romeri, S. Monteil, J. Orloff and A. M. Teixeira, JHEP 1504 (2015) 051.
\bibitem{sterile4}
V. De Romeri, M. J. Herrero, X. Marcano and F. Scarcella,Phys. Rev. D 95(2017)075028.
\bibitem{Zp1}
C. W. Chiang, Y. F. Lin and J. Tandean, JHEP 1111 (2011) 083 .
\bibitem{Zp2}
C. X. Yue and J. R. Zhou, Phys. Rev. D 93(2016)035021.

\bibitem{LR1}
R.N. Mohapatra and J.C. Pati, Phys. Rev. D11(1975)566;
{\it ibid.}{\bf11}(1975)2558..
\bibitem{LR2}
G. Senjanovic and R.N. Mohapatra,Phys. Rev. D12(1975)1502.
\bibitem{sun}
K.-S. Sun, T.-F. Feng, T.-J. Gao and S.-M. Zhao, Nucl. Phys. B 865(2012)486.
\bibitem{dong}
X.-X. Dong, S.-M. Zhao, J.-J. Feng, G.-Z. Ning, J.-B. Chen, H.-B. Zhang, T.-F. Feng, Phys.Rev. D97(2018)056027.

\bibitem{Zp3}
A. Crivellin et al., Phys. Rev.D92(2015)054013.

\bibitem{susy}
R. A. Diaz, R. Martinez, and C. E. Sandoval, Eur. Phys. J. C46 (2006) 403.
\bibitem{HDn}
A. Ilakovac, Phys. Rev. D62(2000)036010.
\bibitem{Lq}
I. de Medeiros Varzielas and G. Hiller, JHEP 06 (2015) 072.
\bibitem{Lq1}
A.D. Smirnov, Mod.Phys.Lett. A33 (2018) 1850019.



\bibitem{PS}
J. C. Pati and A. Salam,  Phys. Rev. D10(1974)275; Erratum-ibid. D11(1975)703.

\bibitem{PDG}
M. Tanabashi et al. (Particle Data Group), Phys. Rev. D 98(2018)030001.
\bibitem{lhcb1}
LHCb Collaboration, Roel Aaij (CERN) et al., JHEP 1803 (2018) 078.
\bibitem{lhcb2}
LHCb collaboration, R. Aaij et al.,Phys.Rev.Lett. 111 (2013) 141801.


\bibitem{Dedes}
A. Dedes, J. Rosiek and P. Tanedo, Phys. Rev. D 79(2009)055006.

\bibitem{X}
Hiren H. Patel, Comput. Phys. Commun. 197(2015)276.
\bibitem{DingY}
G.-J. Ding, M.-L. Yan, Phys. Rev. D 77(2008)014005.
\bibitem{JJZ}
J.-J. Zhang, M. He, X.-G. He, X.-B. Yuan, arXiv:1807.00921.




\end{thebibliography}
\end{document}